\documentclass[twocolumn,showpacs,amsmath,amssymb]{revtex4}
\usepackage{graphicx}

\begin{document}

\title{Gauged Fermionic Q-balls}

\author{Thomas S. Levi$^{1}$ and Marcelo Gleiser$^{2}$}

\address{{\it $^{1}$Department of Physics and Astronomy}
{\it University of Pennsylvania, Philadelphia, PA\quad 19104-6396}\\
{\it $^{2}$Department of Physics and Astronomy}
{\it Dartmouth College, Hanover, NH\quad 03755-3528}\\
}

\begin{abstract}
We present a new model for a non-topological soliton (NTS) that contains
fermions, scalar particles and a gauge field. Using a variational approach,
we estimate the energy of the localized configuration, showing that it can
be the lowest energy state of the system for a wide range of parameters.

\noindent

\end{abstract}

\pacs{11.10.Lm,11.15.Tk}
\maketitle

\section{Introduction}
\label{sec-intro}
The study of solitons and non-topological solitons (NTSs) has a rich history.
They have been proposed as building blocks for stars and black
holes \cite{leepang87,lee87}, and
as dark matter candidates \cite{frieman88,kusenko1,kusenko2}. The first models
for an NTS in 3+1 dimensions were found by Lee and Wick, and by
Friedberg, Lee and Sirlin \cite{leewick74,friedlee76}. The latter NTS contained
one real scalar field to set up a false vacuum in which a second, complex
scalar field was massless, allowing the NTS to be the lowest energy state for
the system under certain conditions.

Coleman and collaborators extended this work to simpler objects dubbed
$Q$-balls, which contained a single complex scalar field that possessed a
conserved global symmetry \cite{coleman85,coleman86}. More recent work has
extended NTSs to contain fermions \cite{macpherson94} and gauge
fields \cite{leestein88}. Finally, work has been done by Kusenko extending
NTSs to some supersymmetric field theories, where the corresponding false
vacuum is set up in the superpotential \cite{kusenko1,kusenko2}.

In this paper we present a new model for an NTS containing mutually interacting fermions, scalar particles, 
and a $U(1)$ gauge field. The fermions have a Yukawa coupling to the scalars, and both
carry a conserved global charge. This model is thus
closer to more realistic theories of particle physics that fit 
in or are inspired by extensions of
the Standard Model. We note that recently a fermionic Q-ball model was proposed by 
Anagnostopoulos et al. \cite{anag}. However, the approach adopted in the present manuscript is
quite different, as we explicitly solve the Dirac equation, as opposed to modeling the 
fermions as a relativistic gas from the outset, and include the Yukawa coupling
to the scalar field. 
Our results, nevertheless, agree with the general conclusions of Ref. \cite{anag},
as we show that, indeed, it is possible to obtain NTSs which are the lowest energy state in the
system for a wide range of parameters. In addition, there is an
opportunity to extend our work to bring it in line with the nuclear bag model, as the 
expression derived below for the energy approximates
some models used in nuclear physics. Throughout we work in natural units where $\hbar = c =1$.

\section{The NTS}
\label{sec-my-stuff}
Consider the Lagrangian

\begin{eqnarray}
\label{my langrangian} {\cal L} = (D_\mu \phi) (D^\mu \phi)^\ast -
\frac{1}{4} F^{\mu \nu} F_{\mu \nu} - U(|\phi |) + \\ \nonumber
 i \overline{\psi} \gamma ^ \mu D_\mu \psi - m
\biggl(1- \frac{|\phi|}{F_-} \biggr) \overline{\psi} \psi ,
\end{eqnarray}
where $\phi$ is the complex scalar field, $D_\mu = \partial _\mu - i e A_\mu$
is the $U(1)$ covariant derivative, $F_{\mu \nu}$ is the field tensor,
and $U(|\phi|)$ is the potential for  the scalar field.
$\psi$ is the 4-component spinor, $\overline{\psi}$ is the Dirac adjoint
spinor, $\gamma ^\mu$ are the four covariant Dirac matrices, $m$ is a
positive constant, and $F_-$ is a constant chosen such that when $|\phi|=F_-$
the fermions are massless within the NTS. We can express the complex
scalar field as two real fields using $\phi = {f\over {\sqrt{2}}} \exp ( i \theta)$,
to get
\begin{eqnarray}
\label{my lagrangian 2}
{\cal L} = \frac{1}{2} \partial _\mu f \partial ^\mu f + \frac{1}{2} f^2
(\partial _\mu \theta - e A_\mu)^2 - U(f)
- \frac{1}{4} F_{\mu \nu} F^{\mu \nu} \\ \nonumber
+ i \overline{\psi} \gamma ^\mu (\partial _\mu - i e A_\mu) \psi -
m \biggl( 1 - \frac{f}{F_-} \biggr) \overline{\psi} \psi .
\end{eqnarray}
We assume the ground state will be spherically symmetric, and will have
no magnetic field and hence no electric currents. Therefore, we may choose a
gauge where $A_\mu = A_0 (r)$. The boundary condition is that
$A_0 (r)\rightarrow 0$
as $r\rightarrow \infty$. In addition, we make the assumption that the
scalar field oscillates in time with a regular frequency and hence
$\theta = \omega t$, where $\omega$ is a positive constant \cite{leestein88}.
The Lagrangian then becomes
\begin{eqnarray}
\label{my total lagrangian} 
L = 4 \pi \int r^2 dr
\biggl [ - \frac{1}{2} f^{\prime 2} + \frac{1}{2e^2} g^{\prime 2} +\frac{1}{2} f^2 g^2 - U(f) \\ \nonumber
+\overline{\psi} \gamma ^0 (\omega - g) \psi - m \biggl (1- \frac{f}{F_-} \biggr ) 
\overline{\psi} \psi + i \overline {\psi} \gamma^\mu
\partial _\mu \psi \biggr ] ,
\end{eqnarray}
where $g \equiv \omega - e A_0 (r)$.
The Euler-Lagrange equations for $g$, $f$ and $\psi$ are
\begin{equation}
\label{my g equation} g'' + \frac{2}{r} g' + \biggl [ e^2 \overline{\psi} \gamma ^0 \psi - e^2 f^2 g \biggr] = 0 ,
\end{equation}
\begin{equation}
\label{my f equation}
f'' + \frac{2}{r} f' + fg^2 - \frac{dU(f)}{df} +
\frac{1}{F_-} m \overline{\psi} \psi = 0 ,
\end{equation}
\begin{equation}
\label{my psi equation}
i \gamma^ \mu (\partial _\mu -  i e A_\mu) \psi - m
\biggl(1 - \frac{f}{F_-} \biggr) \psi = 0 ,
\end{equation}
where it is understood that the only non-vanishing component
of $A_\mu$ is $A_0$. The conserved currents and charges are given by
\begin{equation}
\label{my conserved scalar current}
{\cal J}_\mu ^{scalar} = - i (\phi ^\ast D_\mu \phi - \phi D_\mu \phi ^\ast) ,
\end{equation}
\begin{equation}
\label{my conserved q}
Q = \int {\cal J}_0 ^{scalar} d^3 x = 4 \pi \int r^2 dr f^2 g ,
\end{equation}
\begin{equation}
\label{my conserved spinor current}
{\cal J}_\mu ^{fermion} = \overline{\psi} \gamma _\mu \psi ,
\end{equation}
\begin{equation}
\label{my conserved n}
N = \int {\cal J}_0 ^{fermion} d^3 x = 4 \pi \int r^2 dr \psi ^\dag \psi .
\end{equation}
\section{Taking Care of the Fermions}
To proceed, we could take one of two approaches. We could use the Fermi gas approach as in \cite{macpherson94,anag}. 
Or we could attempt to solve
the equations of motion directly. We attempt the second method here. We begin with the Dirac 
equation for the fermion field. We can write
$\psi$ in terms of two, 2-component, spin-$1/2$ spinors in the chiral representation as
\begin{equation}
\label{2 spinor breakup}
\psi = \left( \begin{array}{l}
\phi _R \\
\phi _L
\end{array}
\right) ,
\end{equation}
where $\phi _{R(L)}$ is a right (left)-handed spinor. It is easier to proceed
if we switch to the non-covariant representation of the Dirac equation.
Multiplying by $\beta ^{-1} \equiv (\gamma ^0)^{-1}$ on the left and using
that $\gamma ^i \equiv \beta \alpha ^i$, we get the equation
\begin{equation}
\label{time-dependent dirac equation}
i \frac{\partial \psi}{\partial t} = (-i {\bf \alpha} \cdot \nabla +
\beta M +V_f (r) )\psi ,
\end{equation}
where $V_f (r) = e A_0 (r)$ is the potential for the fermions, and
$M = (1 - f/ F_-)m$ is the mass of the fermion. Following Lee et al.\,
we will assume that $f(r) = constant = F_- \equiv F$ inside the NTS
\cite{leestein88}. We will also assume that $\psi$ is of the simple
form $\psi (r,t) = \psi (r) \exp(-i E t)$ where $E$ is the energy of a
single fermion. We therefore get the new equation
\begin{equation}
\label{second dirac equation}
E \psi = (-i {\bf \alpha} \cdot \nabla +V_f (r) ) \psi .
\end{equation}
We desire $\psi$ to be spherically symmetric in the ground state.
Since $\phi _R$ and $\phi _L$ must have opposite parity under spatial
reflection, they cannot both be symmetric. Therefore, we must choose
one of them to equal zero, and conventions of a right-handed
coordinate system dictate that it is the left handed component that must be zero.
We can now expand in terms of an angular part and a
radial part by writing \cite{greiner}
\begin{equation}
\label{separation of variables for the spinor}
\phi_R = i h(r) \Omega_{jlm} ,
\end{equation}
where $\Omega_{jlm}$ is a spherical spinor and the indices $j,l,m$ are the
quantum numbers of total angular momentum, orbital angular momentum, and
the $z$-component of angular momentum, respectively. In the
spherically-symmetric ground state we have $l=0$ and so $j=l+1/2=1/2$, and
the spherical
spinor for this case simplifies to \cite{greiner}
\begin{equation}
\label{ground state spinor}
\Omega _{\frac{1}{2} 0 \frac{1}{2}} = \left(
\begin{array}{c}
Y_{00} \\
0 \end{array} \right) = \sqrt{\frac{1}{4 \pi}} \left( \begin{array}{c}
1 \\
0
\end{array} \right) ,
\end{equation}
and we see that indeed our wave-function is spherically
symmetric. The equations for $h(r)$ become \cite{greiner}
\begin{equation}
\label{h equation 1}
\frac{d h}{dr} + (1 + \kappa) \frac{h}{r} = 0 ,
\end{equation}
\begin{equation}
\label{h equation 2}
[E - V_f (r) ] h = 0 ,
\end{equation}
where $\kappa \equiv -(l+1)$ so here $\kappa = -1$ \cite{greiner}. The
solution to Eq.\ (\ref{h equation 1}) is then simply $h=constant \equiv B$.
However, we notice a problem with Eq.\ (\ref{h equation 2}); namely, if
$V_f (r)$ is non-constant then it demands that $h=0$ since $E$ must be a
constant, and the equation must be sastified for all $r$. How are we to get
around this problem? We can approximate the fermionic energy inside the NTS as constant
by taking the expectation value of the fermionic potential inside it.
Then we can write $E \approx \langle E \rangle = \langle V_f (r) \rangle$ and
Eq.\ (\ref{h equation 2}) may be approximately satisfied inside the NTS. Before doing this we note that if
$h$ is constant, the expression for the conserved charge
Eq.\ (\ref{my conserved n}) becomes simply
\begin{equation}
\label{simplified fermion charge}
N = \frac{B^2 R^3}{3} ,
\end{equation}
where $R$ is the radius of the ball. We then get that
\begin{equation}
\label{average v}
\langle V_f \rangle = 4 \pi \int _0 ^R \psi ^\dag \sigma _3 V(r) \psi r^2 dr =
\frac{3N}{R^3} \int _0 ^R r^2 [\omega - g(r)] dr ,
\end{equation}
where $\sigma _3$ is the third Pauli spin matrix, and our approximate
solution is
\begin{equation}
\label{final solution to dirac equation}
\psi (r,t) = \frac{1}{\sqrt{8 \pi}} \biggl( \frac{3N}{R^3} \biggr)^{1/2}
\exp \biggl( -it \langle V_f \rangle \biggr)
\theta (R-r) \left( \begin{array}{c}
1 \\
1 \\
0 \\
0
\end{array} \right) ,
\end{equation}
where $\theta (R-r)$ is the step function.

We can now use Eq.\ (\ref{final solution to dirac equation}) in Eq.\ (\ref{my g equation}). The solution for $g(r)$ is
\begin{equation}
\label{g solution}
g(r) = \left\{ \begin{array}{l}
[\omega - e^2 Q / 4 \pi R] R \sinh(eFr)/r\sinh(eFR) + \\
 \psi^{\dagger} \psi/F^2, \ \ \ r \leq R \\
\omega - e^2 Q / 4 \pi r, \ \ \ r>R ,
\end{array}
\right.
\end{equation}
where $R$ is the radius of the soliton. We can now substitute for $f^2 g$ in Eq.\ (\ref{my conserved q}) to get
\begin{equation}
\label{asymptotic form of charge} e^2 Q = \int _0 ^R 4 \pi dr [( g' r^2)' + e^2 r^2 \psi ^{\dagger} \psi ].
\end{equation}
Using our solution for $g(r)$, Eq.\ (\ref{g solution}), we get
\begin{equation}
\label{omega q equation} \omega= \frac{e^2 Q}{4 \pi R} \biggl[ \frac{x}{x-\tanh(x)} \biggr] - 
\frac{e^2 N \tanh(x)}{4 \pi R (x -\tanh(x))},
\end{equation}
where $x\equiv eFR$.

\section{Estimating the Gauged Fermionic Q-Ball Energy}
The gauge invariant energy can be written as
\begin{eqnarray}
\label{energy} E = 4 \pi \int r^2 dr \biggl[ \frac{1}{2} f^{\prime 2} + \frac{1}{2 e^2} 
g^{\prime 2} +\frac{1}{2} f^2 g^2 +\\ \nonumber 
U(f) +i\overline{\psi} \gamma ^i \partial _i \psi \biggr] + E_F~,
\end{eqnarray}
where $E_F$ is the relativistic Fermi energy given by $E_F = (3 \pi /4)(3/2
\pi)^{2/3} (N^{4/3} / \Delta)$. Ignoring surface terms ${\cal O}(R^2)$,
performing some partial integrations, and using our solution for
$\psi$ we get, for $r \leq R$,
\begin{equation}
\label{second form of approximate energy} E \leq \frac{1}{2} \omega Q + \frac{4}{3} \pi R^3 U(F) + E_F
\end{equation}
We can use Eq.\ (\ref{omega q equation}) to get
\begin{eqnarray}
\label{my integrated out energy} E \leq \frac{e^2 Q^2}{8 \pi R} \biggl[ \frac{x}{x-\tanh x} \biggr] + 
\frac{4}{3} \pi R^3 U(F) +
\\ \nonumber  +\frac{C_1 N^{4/3}}{R} - \frac{e^2 Q N}{8 \pi R} \biggl[ \frac{\tanh x}{ x - \tanh x}\biggr],
\end{eqnarray}
where $C_1 \equiv 3 \pi /4 (3/2 \pi)^{2/3}$. The next step is to minimize this expression with respect to the 
various parameters at fixed
$N$ and $Q$. We examine the case of small $e$ and expand to order $e^3$, where the Laurent series are given 
by $x/(x-\tanh x) \approx 3/x^2
+ 6/5$ and $\tanh x / (x-\tanh x) \approx 3/x^2 + 1/5$. We then get an approximate form for $E$ to 
${\cal O} (e^3)$ of
\begin{eqnarray}
\label{my series expanded energy} E \leq \frac{3 Q^2}{8 \pi F^2 R^3} + \frac{4}{3} \pi R^3 U(F) + \frac{C_1 N^{4/3}}{R} + \\
\nonumber \frac{3 e^2 Q^2}{20 \pi R} -\frac{3QN}{8 \pi F^2 R^3} - \frac{e^2 Q N}{40 \pi R} .
\end{eqnarray}
These terms now have an easy physical interpretation. The first term is the zero-point energy of the scalar 
particles. The second term is
the vacuum volume energy of the bag. The third term is the Fermi energy. The fourth term is the Coulomb 
repulsion of the scalar particles.
The fifth and six terms represent the interactions of the fermions with the scalars and gauge fields, 
which may significantly alter the NTS's energy.

\begin{figure}
\includegraphics[width=245pt,height=220pt]{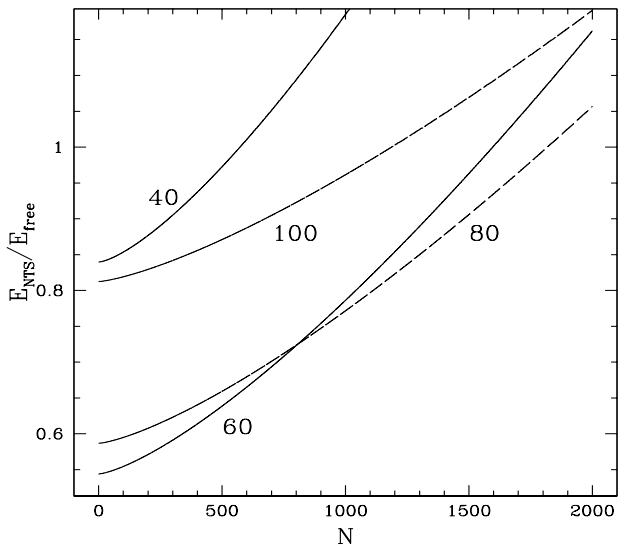}
\caption{ \label{en_n}
Ratio of NTS energy to the energy of free fermions and scalars as a function of fermionic charge $N$. The labels
indicate the different values of the NTS radius.
}
\end{figure}

If we minimize this expression with respect to $R$ we obtain, writing $R^2 \equiv y$,
\begin{eqnarray}
\label{y equation} 4 \pi U(F) y^3 - y \biggl( \frac{3e^2Q^2}{20 \pi} + C_1 N^{4/3} - 
\frac{e^2 Q N}{40 \pi} \biggr) - \\ \nonumber
\frac{9Q}{8 \pi F^2} (Q-N) = 0.
\end{eqnarray}
The formal solution to this equation is given by
\begin{eqnarray}
\label{formal y solution}
z=\cos \biggl[ \frac{1}{3} \arccos \biggl( \frac{3B}{2A} \sqrt{\frac{3}{A}} \biggr) + \frac{2 \pi n}{3} \biggr] , \\
n=0, \pm 1, \pm 2 \ldots \nonumber \\  z \equiv y/ \alpha , \ \ \ \ \alpha \equiv \sqrt{ (4 A /3) }, \nonumber \\  
A=\frac{C_1 N^{4/3}}{4
\pi
U(F)} + \frac{e^2 Q}{160 \pi^2 U(F)} (6Q-N), \\
B= \frac{9Q}{32 \pi ^2 F^2 U(F)} (Q-N) .
\end{eqnarray}
Unfortunately, since we must minimize the energy with respect to both $R$ and $F$ the details of selecting 
the correct root (or even if a
positive real root exists) depend on the nature of $U(F)$. Hence, it is not possible to write down a general 
procedure for selecting a root.
Once a suitable potential has been chosen however, the solution follows in a straightforward manner. 
To illustrate this, we solve for the
potential
\begin{equation}
\label{phi 6 potential} U(f) = \frac{\lambda ^2 f^6}{6 \mu ^2} - \frac{f^4}{4} + \frac{\mu ^2 f^2}{2} .
\end{equation}
We choose $\lambda=0.444$, $\mu =.25$, $e=.1$, $Q=10000$, and $M_{\psi} =2$ (we must choose $\lambda ^2 > 3/16$ to 
insure that $U(f)
> 0$ for all $f \neq 0$), where $\mu$ and $M_{\psi}$ have dimensions of mass ($M_{\psi}$ is the mass of the 
fermions in the true vacuum).
The energy scale is set by $F$, which has dimensions of mass. We first solve the problem with no fermions 
present, which is identical to the
scenario considered in \cite{leestein88}. This gives us an idea of a possible range of values for $R$ 
and $F$ (based on this analysis, we set $F=0.48$ throughout) to
search for fermionic NTSs. Writing the energy of the free scalars and fermions as
$E_{\rm free} = \mu Q + M _\psi N$, the NTS is stable whenever $E_{NTS} / E_{\rm free} < 1$. 
In Fig. \ref{en_n},  we show the ratio $E_{NTS} / E_{\rm free}$ as a function of
increasing fermion number, for several values of the radius. From a more detailed analysis we can show that,
for the parameters used in Fig. \ref{en_n},
the condition for the existence of NTSs is satisfied for $35 \lesssim R \lesssim 111$. Clearly, the same sort of
range search can be performed for any set of paramenters.

\begin{figure}
\includegraphics[width=245pt,height=220pt]{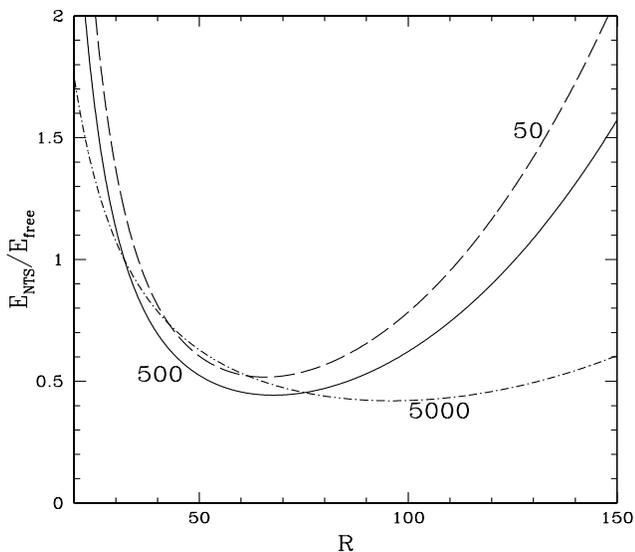}
\caption{ \label{en_r}
Ratio of NTS energy to the energy of free fermions and scalars as a function of radius $R$. The labels
indicate the different values of the fermionic charge $N$.
}
\end{figure}

In Fig. \ref{en_r} we show the ratio $E_{NTS} / E_{\rm free} $ as a function of the NTS radius for different
values of the fermionic charge $N$. Clearly, there is a wide range of values for $N$ wherein the NTS is
the preferred energy configuration. For large radii (larger than the range of values displayed in the Figure), 
the NTS energy is independent of $N$, as can be easily seen
from Eq. (\ref{my series expanded energy}).

We conclude with a few remarks about our NTS. First, we see that if we use a potential that satisfies 
Coleman's condition \cite{coleman85}
$\min [2U/|\phi ^2|] < \mu ^2$ ($\mu$ is the mass of the free scalar particles), in order to set up a 
false vacuum where the fermions
can be massless, then we can always find parameters where $E_{NTS} < E_{\rm free}$. 
Second, we see that the presence of both fermions 
and the gauge field increases
the energy and the radius of the NTS, while the attractive Yukawa coupling between the fermions and 
scalars decreases both in relation
to the ungauged scalar $Q$-balls studied by Coleman and collaborators \cite{coleman85, coleman86}. Third, 
we note that the asymptotic form
of the energy is
\begin{equation}
\label{asymptotic energy} \lim_{R  \rightarrow \infty} E \rightarrow \frac{4}{3} \pi R^3 U(F) .
\end{equation}
Hence, if we scale $U(F)$ in such a way that $R^3 U(F) = constant < \frac {3}{4} E _{\rm free}$, 
our NTS can be stable at arbitrarily large
radii. Physically, we have obtained a state of matter with scalar particles and fermions uniform throughout. 
We thus see that it is possible to form a
NTS out of gauged fermions and scalar particles, which can be the preferred energy state of the system for a 
wide range of parameters. It would be interesting to investigate the solutions to the set of coupled equations
numerically, obtaining a more detailed analysis of the allowed parameter space.

\acknowledgments We thank Robert Caldwell and Walter Lawrence for offering valuable advice and criticism. We also
thank N. Tetradis for alerting us to the work of Ref. \cite{anag} and to a problem in the original manuscript.
We also thank the referee for his/hers remarks, which forced us to reconsider our original assumption concerning
the solution of the coupled fermionic-scalar-gauge field equations.
MG thanks the ``Mr. Tompkins Fund
for Cosmology and Field Theory'' at Dartmouth, and NSF -- grants PHY-0070554 and PHY-0099543 for partial 
financial support.

\end{document}